\documentclass[useAMS,usenatbib]{mn2e}
\usepackage{graphicx}
\def\jcoph{J. Comp.\ Phys.}

\def\eg{{\it e.g.}}
\def\etal{{\it et al.}}

\def\ie{{\it i.e.}}

\def\pmb#1{\setbox0=\hbox{$#1$}%
  \kern-0.25em\copy0\kern-\wd0
  \kern.05em\copy0\kern-\wd0
  \kern-0.025em\raise.0433em\box0}

\def\GCS{{\small GCS}}
\def\Hipp{{\it Hipparcos}}
\def\RAVE{{\small RAVE}}
\def\SDSS{{\small SDSS}}
\def\SSPP{{\small SSPP}}
\def\LSR{{\small LSR}}
\def\ILR{{\small ILR}}
\def\OLR{{\small OLR}}

\title[Velocity-space substructure]{Velocity-space substructure
from nearby RAVE and SDSS stars}
\author[Hahn, Sellwood and Pryor]{Chang Hoon Hahn$^{1}$\thanks{E-mail:
chahah@eden.rutgers.edu}, J. A. Sellwood$^{1}$\thanks{E-mail:
sellwood@physics.rutgers.edu} and Carlton Pryor$^{1}$\thanks{E-mail:
pryor@physics.rutgers.edu} \\
$^{1}$Rutgers University, Department of Physics \& Astronomy, 136
Frelinghuysen Road, Piscataway, NJ 08854-8019, USA}

\begin{document}

\date{Accepted 2011 Aug 15. Received 2011 July 16; in original form 2011 January 17}

\pagerange{\pageref{firstpage}--\pageref{lastpage}} \pubyear{2010}

\maketitle

\label{firstpage}

\begin{abstract}
We extract a sample of disc stars within 200~pc of the Sun from the
\RAVE\ and \SDSS\ surveys.  Distances are estimated photometrically
and proper motions are from ground-based data.  We show that the
velocity-space substructure first revealed in the Geneva-Copenhagen
sample is also present in this completely independent sample.  We also
evaluate action-angle variables for these stars and show that the
Hyades stream stars in these data are again characteristic of having
been scattered at a Lindblad resonance.  Unfortunately, analysis of
such local samples can determine neither whether it is an inner or an
outer Lindblad resonance, nor the multiplicity of the pattern.
\end{abstract}


\section{The Local Distribution of Stellar Velocities}
Analysis of the \Hipp\ data by \cite{Dehn98} revealed that the stellar
velocity distribution in the solar neighbourhood manifested
significant substructure.  The heroic Geneva-Copenhagen Survey
\citep[hereafter GCS][]{Nord04,HNA9} followed up with radial velocity
measurements of 14\,139 nearby F and G dwarf stars, which confirmed
and strengthened Dehnen's conclusion.

The Geneva-Copenhagen survey was constructed so as to avoid most of
the selection biases that went into the full \Hipp\ sample.  Aside
from a concentration of 112 stars in the Hyades cluster, the
distribution of sample stars over the sky is remarkably uniform, with
a slightly higher density in the declination range south of $\delta =
-26^\circ$.  The large majority of stars are within 200~pc of the Sun
and the sample within 40~pc is believed to be nearly complete.

The Cartesian heliocentric velocity components of stars near the Sun
in Galactic coordinates are $U$, $V$ and $W$, with $U$ being directed
towards the Galactic centre, $V$ being in the direction of Galactic
rotation, and $W$ towards the north Galactic pole.  The velocity
substructure in the \GCS\ is particularly evident in the $U-V$ plane,
where the distribution is broken into a number of substantial streams,
with no underlying smooth component.  Numerous studies
\citep{Fama07,Bens07,BH10,Pomp11} have shown that the streams are both
too substantial and chemically inhomogeneous to be dissolved star
clusters \citep[\eg][]{Egge96}.

Dynamical evolution is the most likely source of the features, which
have been modelled extensively.  \cite{dSWT} suggest that the entire
velocity distribution arises from a succession of transient spiral
perturbations, while \cite{Helm06} attribute some substructure to
minor accretion events.  Individual features have been modelled as
responses to the bar and/or various assumed spiral perturbations
within the disc \citep{Dehn00,Quil03,QM05,Chak07,Anto09,MBSB10}.  The
analysis of \citet{QM05}, which was extended by \cite{Pomp11},
identified the inner ultra-harmonic resonance of an assumed 2-arm
spiral as a possible cause of both the Hyades and Sirius streams.

By contrast, \citet[][hereafter Paper I]{Sell10} presented an analysis
that did not need to assume a form for the perturbation.  Using
action-angle coordinates, he showed that the stars of the Hyades
stream were both concentrated along a resonance line in action space
and grouped in a combination of angle coordinates (possibly)
indicative of a recent inner Lindblad resonance (hereafter \ILR).
\citet{McM11} has confirmed most of his analysis, and concurs that a
Lindblad resonance is responsible but, unfortunately, was able to show
that subtle selection effects in such local data imply that the
distribution could also be consistent with trapping in an outer
Lindblad resonance (\OLR).  We discuss selection effects and
McMillan's analysis in the Appendix.

While the \GCS\ sample offered compelling evidence for a Lindblad
resonance, it is desirable to attempt to confirm the result from other
independent surveys.  The Radial Velocity Experiment
\citep[\RAVE,][]{Stei06}, a large spectroscopic survey of stars in the
southern sky, plans to measure the heliocentric radial velocities and
stellar parameters for about a million stars in the apparent magnitude
range $9 < m_I < 12$; the first $21\,121$ were made available in the
second data release \citep{Zwit08}.  The typical uncertainty in the
radial velocity is $<2\;$km~s$^{-1}$, but the distance to most stars
has to be judged photometrically and most proper motions are from
ground-based data.  Thus the three phase space coordinates for each
star are of much lower quality than are those in the Geneva-Copenhagen
survey, although this weakness will, when the survey is complete, be
compensated by a much larger sample size.  The huge northern Sloan
Digital Sky Survey \citep[\SDSS,][]{York00} and Segue2 \citep{Yann09}
surveys are complete, but sample fainter stars (the magnitude range
for Segue2 was $14.0 < g < 20.3$) that are therefore generally more
distant than are the \RAVE\ stars.  The recently-released M-dwarf
sample of \SDSS\ stars \citep{West11} substantially increases the
number of stars with estimated distances and kinematics within the
neighbourhood of the Sun.

\section{Sample selection}
\subsection{RAVE stars}
We have downloaded the on-line table of the second data release from
the \RAVE\ website and selected a subset of stars for
analysis.\footnote{The third release occurred while this paper was
  under review.}  We estimate distances to these stars by fitting to
the Yonsei-Yale isochrones \citep{Dema04} using a method related to
that described by \cite{Bred10}.  We adopt many of their selection
criteria: we require the spectral signal-to-noise parameter $>20$ with
a blank spectral warning flag field; the parameters [M/H], log($g$),
and $T_{\rm eff}$ to be determined; and the stars to have J \& K$_s$
magnitudes from two-micron all sky survey \citep[2MASS][]{Skru06} with
no warning flags about the identification of the star or the 2MASS
photometry.  Unlike those authors, however, we have kept stars with
$b<25\degr$ on the grounds that extinction for the nearby stars that
interest us will not be large enough in the near IR to severely bias
our distance estimates.  As we wish to select nearby main-sequence
stars that are members of the disc population, we also eliminate stars
with log$(g)<4$, $T_{\rm eff}> 10^4\;$K, and with
$|v_r|>80\;$km~s$^{-1}$.

\begin{figure}
\begin{center}
\includegraphics[width=.9\hsize,angle=0]{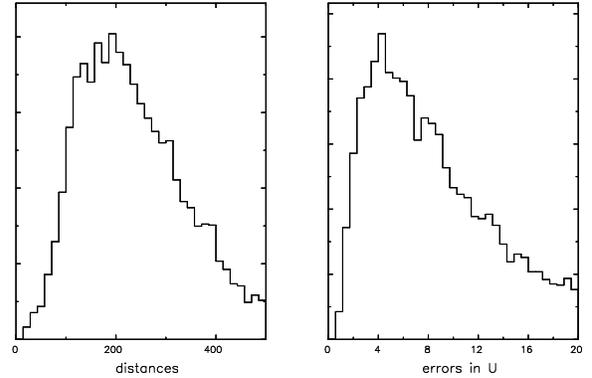}
\end{center}
\caption{Histograms of distances and $U$-velocity uncertainties for the
  selected sample of 5145 \RAVE\ stars.}
\label{plotrave}
\end{figure}

We estimate the absolute J magnitude of each selected star by matching
the estimated [Fe/H], log($g$), $T_{\rm eff}$, and \hbox{J-K$_s$}
colour to values in the isochrone tables for stars of all ages and all
values of [$\alpha$/Fe], rejecting a few more stars for which the best
match $\chi^2>6$.  We consider the closest match in the tables to the
given input parameters to yield the best estimate of the absolute
magnitude from which we estimate a photometric distance using the
apparent J-band magnitude.\footnote{\citet{Zwit10} describe a similar
  method, but define a ``most likely'' estimate of the absolute
  magnitude that differs from our ``best'' estimate.  The difference
  is likely to be well within the uncertainties for the main-sequence
  stars considered here.}  Our resulting sample contains $7\,384$
stars.  We save the values of [Fe/H], log($g$), $T_{\rm eff}$, and
J-K$_s$ colour of the closest matching model star in the isochrone
table; Monte-Carlo variation of the stellar parameters about this
saved set of values suggests that distances have a relative precision
of 30\% -- 50\%, with some larger uncertainties.

We use the proper motions in equatorial coordinates tabulated in
\RAVE, mostly from Tycho-2 \citep{Tycho}, which we then combine with
the radial velocity and position to determine the heliocentric
velocity in Galactic components $U, \; V\; \& \;W$
\citep{JS87,Piatek}.  We estimate uncertainties in these velocities
from 5000 Monte Carlo re-selections of all the stellar parameters that
affect the distance estimate, adopting $\sigma({\rm J}) = 0.03\;$mag,
$\sigma({\rm J-K}_s) = 0.042\;$mag, $\sigma(T_{\rm eff})=300\;$K,
$\sigma(\log\,g)=0.3\;$dex, and $\sigma([{\rm Fe/H}])=0.25\;$dex
\citep{Bred10}, as well as the tabulated radial velocity and proper
motion uncertainties.

In order to select nearby thin disc stars, we further restrict the
sample to stars whose best estimate of the distance is within
$500\;$pc and retain only those having an energy of vertical motion
about the Galactic mid-plane, $E_z = 0.5(z^2\nu^2 + W^2) <
392\;$(km~s$^{-1})^2$, with the vertical frequency $\nu =
0.07\;$km~s$^{-1}$~pc$^{-1}$ \citep{BT08}, giving them a maximum
vertical excursion of $\pm 400\;$pc.  This latter restriction rejects
most thick disk and halo stars, as well as a few thin disc stars.

Fig.~\ref{plotrave} shows histograms of distances and of velocity
uncertainties for the $5,145$ remaining stars.  The mode of the
distance distribution is 200~pc from the Sun.  The estimated
uncertainties in $U$ are generally $< 10\;$km~s$^{-1}$, with a tail up
to values four times larger.

\begin{figure}
\includegraphics[width=1.3\hsize,angle=270]{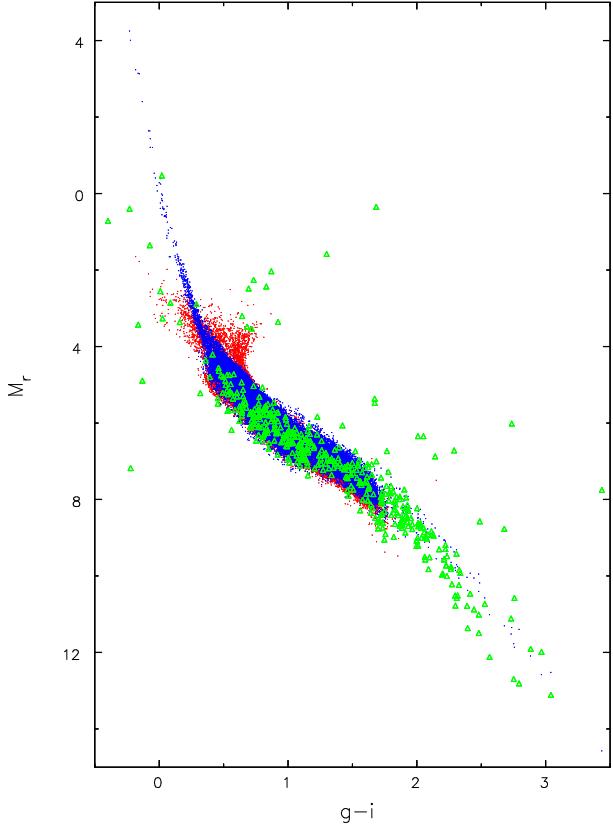}
\caption{Blue points show the estimated M$_r$ magnitude as a function
  of colour and metallicity using the formulae from \citet{Ivezic}.
  Red points show the same quantity obtained from isochrone fitting,
  while green triangles show values for which no acceptable isochrone
  fits the star. See text for discussion.}
\label{ivezic}
\end{figure}

\subsection{SDSS and Segue2 stars}
We also selected stars from the DR7 of \SDSS\ \citep{Abaz09} with
$4<\log(g)<5$, $3000<T_{\rm eff}< 10^4\;$K, and with
$|v_r|<60\;$km~s$^{-1}$, as estimated by the Sloan spectral parameters
pipeline \citep[][\SSPP]{Pipe}, which yielded over $100\,000$
candidate stars.  As for \RAVE, we wish to estimate photometric
distances to these stars.

\cite{Juric} and \cite{Ivezic}, respectively, propose formulae for
estimating the absolute magnitudes, $M_r$, from the $r-i$ colour, or
from the $g-i$ colour and [Fe/H] but, as for the \RAVE\ stars, we have
preferred to estimate distances by fitting the full stellar parameters
to isochrone tables.  The Dartmouth isochrone tables \citep{Dotter}
give absolute magnitudes in Sloan colour bands.  We interpolate for
[Fe/H], for all ages and with [$\alpha$/Fe]=0 only, to estimate $M_i$
by fitting $r-i$ colours, together with \SSPP\ estimates of [Fe/H],
log$(g)$, and $T_{\rm eff}$.  As recommended by \cite{Pipe}, we
increased the uncertainties in the estimated stellar parameters to
$\sigma(T_{\rm eff})=157\;$K, $\sigma(\log\,g)=0.28\;$dex, and
$\sigma([{\rm Fe/H}])=0.24\;$dex.  We find 38\,690 stars whose best
estimate of distance modulus in the $i$-band places them within 2~kpc.

Fig.~\ref{ivezic} illustrates the superiority of isochrone fitting
over relying on a single colour index.  The blue points show the
estimated $M_r$ from the fifth-order polynomial function of the $g-i$
colour, broadened by a spread given by a quadratic expression in
[Fe/H], as recommended by \citet[][eqs.~A2 \& A7]{Ivezic}.  The red
points show the isochrone-fitted $M_r$ magnitude obtained as described
above; note that the \SSPP\ does not return parameters for stars
having $T_{\rm eff}<4\,500$, which is the reason for the absence of
credible main sequence stars redward of $g-i \sim 1.7$.  For the large
majority of stars, the blue and red points fill the same region of the
Figure, and are therefore consistent; the two magnitude estimates
differ by $<0.2$ mag for 89\% of the stars.
But, as is physically reasonable, isochrone fitting broadens the
distribution near the main sequence turn off, although the spread may
be spuriously enhanced by errors in the pipeline parameters,
especially in $\log(g)$.  The principal advantage of isochrone
fitting, however, is that it flags stars with inconsistent parameters.
The green triangles show 938 stars that had $\chi^2>6$, based on the
above uncertainties, for the smallest difference between the estimated
$r-i$ colour, $T_{\rm eff}$, and $\log(g)$ of any model star in the
Dartmouth isochrone tables.  A large $\chi^2$ generally arises for a
star that has a colour (we used $r-i$) estimate inconsistent with the
pipeline-estimated $T_{\rm eff}$, although the estimated $\log(g)$ may
be inconsistent in other cases.  We therefore adopt the absolute
magnitudes given by the isochrone fits to estimate distances,
discarding the stars having a large $\chi^2$.

\begin{figure}
\begin{center}
\includegraphics[width=.9\hsize,angle=0]{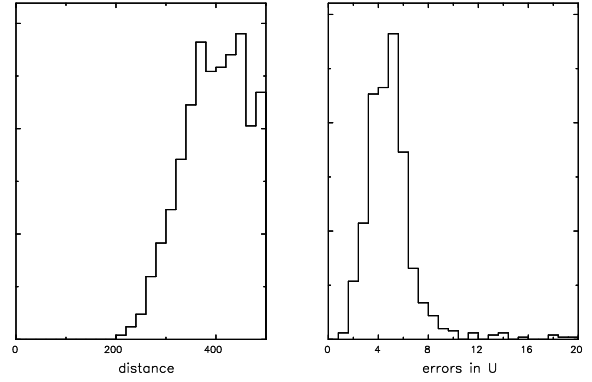}
\end{center}
\caption{Histograms of distances and $U$-velocity uncertainties for the
  selected sample of 629 \SDSS\ stars.}
\label{plotsdss}
\end{figure}

We adopted the distance modulus in $i$, and limited the distance and
$z$ motions as for \RAVE\ stars.  These restrictions reduced the
sample from the \SSPP\ to just 629 nearby disc stars.  This
disappointing number results from the apparent magnitude limit for
\SDSS\ spectroscopy of $m_i \ga 14$ coupled with the elimination of
all M-dwarfs from the sample through the low-temperature restriction
of the \SSPP; the distance histogram shown in Fig.~\ref{plotsdss}
indicates that no star from the \SSPP\ is closer than 200~pc.  Because
uncertainties in the stellar parameters are somewhat smaller than for
\RAVE\ stars, the estimated uncertainties in distance are slightly
smaller, which translates into smaller velocity uncertainties.

\begin{figure}
\begin{center}
\includegraphics[width=.9\hsize,angle=0]{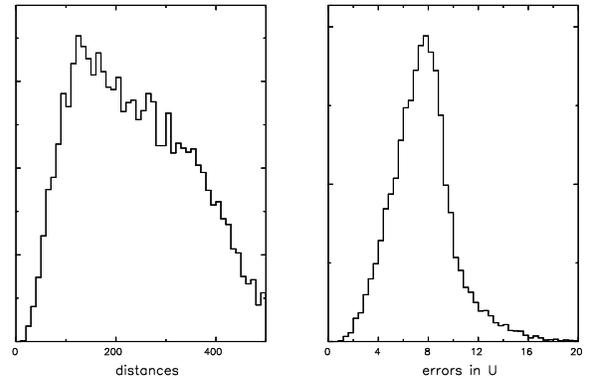}
\end{center}
\caption{Histograms of distances and $U$-velocity uncertainties for the
  selected sample of 10\,669 M-dwarf stars.}
\label{plotMds}
\end{figure}

\setbox1\vbox{
\hbox{\hspace{1.8cm}\includegraphics[width=.9\hsize,angle=0]{skyproj.ps}}
\vspace{.3cm}

\includegraphics[width=1.1\hsize,angle=0]{GCScont.ps}

\vspace{-0.7cm}}

\begin{figure*}
\begin{center}
\hbox{
\includegraphics[width=.44\hsize,angle=0]{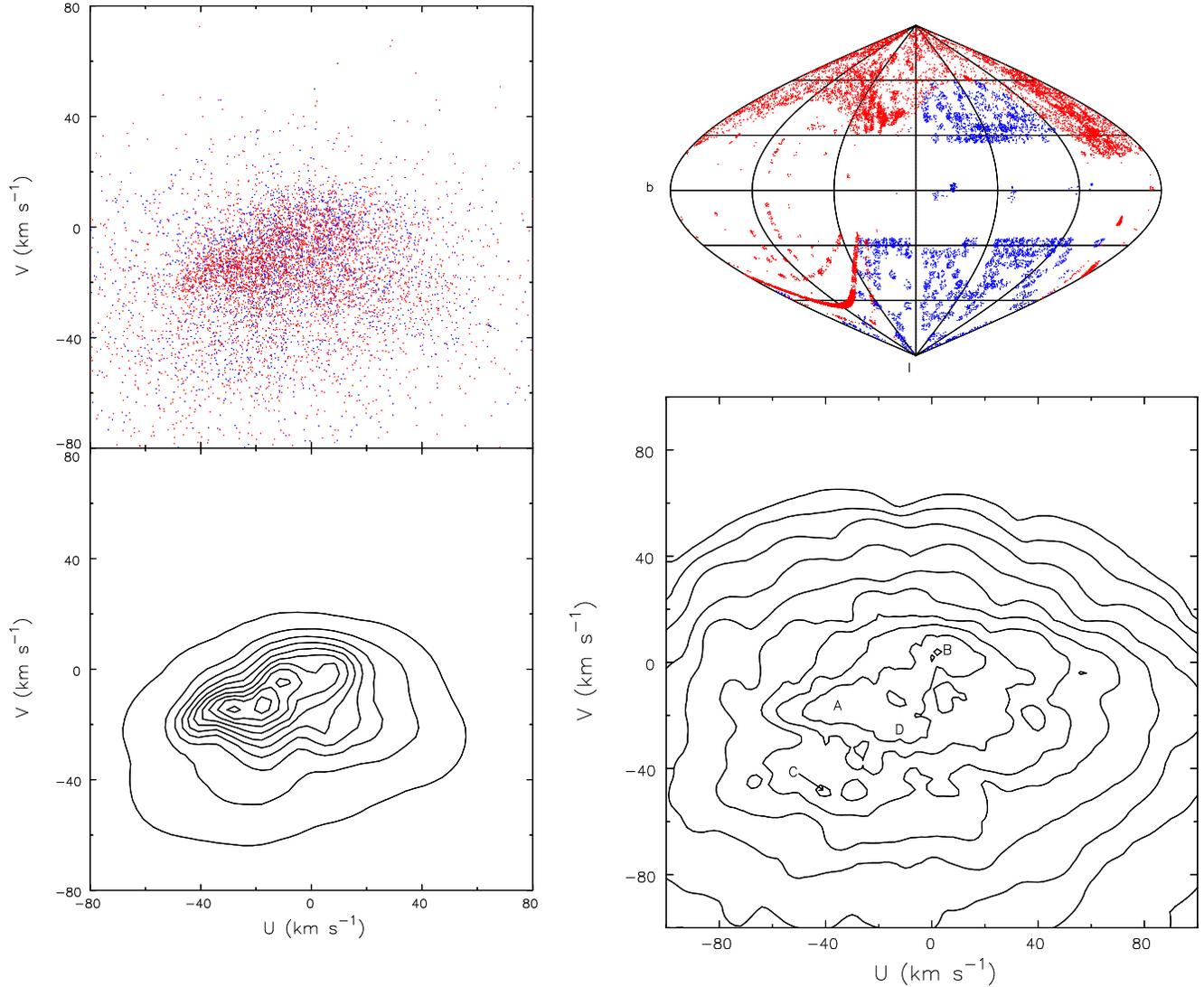} \hspace{-0.4cm}
\box1}
\end{center}
\caption{Top right: the sky distribution in Galactic coordinates of
  the selected \RAVE, blue, and \SDSS\ M-dwarf, red, stars within
  200~pc of the Sun.  Top left: the Galactic velocity components of
  the selected stars with the same colour coding as in the top panel.
  Bottom left: contours of the density of the combined sample in
  velocity space.  The shapes of the contours are quite similar to
  those for the Geneva-Copenhagen sample reproduced from Paper I in
  the bottom right panel, where the letters indicate the approximate
  locations of the principal ``star streams'': A -- Hyades, B --
  Sirius, C -- Hercules, and D -- Pleiades. }
\label{contall}
\end{figure*}

\subsection{M-dwarf sample}
\cite{West11} provided a catalogue of 70\,841 M-dwarfs from \SDSS\ DR7
that were omitted from the \SSPP.  Their table provides a photometric
estimate of the distance to each star, as well as a radial velocity
and proper motion from USNO-B/SDSS catalogue \citep{Munn04,Munn08}.
They suggest that distance uncertainties are typically about 20\% and
uncertainties in radial velocity are 7-10~km s$^{-1}$.

As \cite{West11} recommend, we selected stars with the ``goodPM'' and
``goodPhot'' flags set to `true', and the ``WDM'' flag set to `false'
to eliminate possible binaries with a white dwarf companion, which
reduces the sample to 39\,151 stars.  We further excluded stars having
no radial velocity as well as those with no distance estimate or for
which the estimated distance exceeded 500~pc.  Finally, we also
eliminated stars for which any component of the heliocentric velocity
exceeded 80~km s$^{-1}$ and those having a vertical energy that would
take them farther than 400~pc from the disc mid-plane, leaving us with
a final sample of 10,669 stars.

We estimated uncertainties in the velocity components, $U$, $V$ \& $W$
by combining the 20\% distance uncertainty, a 10~km s$^{-1}$
uncertainty in the radial velocity, and the proper motion
uncertainties.  The distribution of distances and $U$ velocity errors
is shown in Fig.~\ref{plotMds}; many of these intrinsically faint
stars lie within 200~pc and, again, velocity uncertainties are
typically $\sim 10\;$km s$^{-1}$.

\subsection{Combined sample}
The three surveys yielded a total sample of some 16\,443 stars within
500~pc.  Since few stars in these surveys are close to zero Galactic
latitude, the in-plane velocity components $U$ and $V$ are mostly
determined by the proper motion, with the more-precise radial velocity
making a smaller contribution.  For this reason, we confine all the
following analysis to the $6,769$ stars closer than 200~pc to the Sun,
which have correspondingly smaller uncertainties in these two velocity
components.  We refer to this as our final ``combined sample''.

Note that the combined sample contains just four stars from the
\RAVE\ survey, and no stars from the \SDSS, that were also in the
\GCS.  The reason for this tiny overlap is that the \GCS\ was limited
to F and G dwarf stars brighter than $m_{\rm vis}=8.6$, and the only
stars from \SDSS\ in the combined sample are the M-dwarfs.  Thus the
two samples are truly independent; even the radial velocities of
the four stars in common were remeasured by \RAVE.

The top right panel of Fig.~\ref{contall} shows that the combined sample
covers most of the sky, except for the Galactic plane, although far
from uniformly.  The top left panel shows the best fit velocity
components in Galactic coordinates.  Although velocity uncertainties
are several times larger than for the \GCS, the general appearance of
the distribution in velocity space is similar.

The bottom left panel of Fig.~\ref{contall} contours the density in
the space of these two velocity components, using linearly-spaced
contour levels.  The bottom right panel reproduces the plot, with
logarithmically spaced contours, constructed from the \GCS\ sample
that was shown as Fig.~2 in Paper I.  Many of the features in the
$(U,V)$-plane that stand out in the \GCS\ have counterparts in this
completely independent sample, although their velocity locations do
not match perfectly.  The strongest feature is the Hyades stream, but
the bottom left panel has clear hints of the Pleiades, Sirius, and
even Hercules streams.  Note that errors in our distance estimates
will give rise to correlated errors in the $U$ and $V$ velocity
components, since they are derived mostly from proper motions.
However, the misplacements of points in the $(U,V)$-plane are in
directions that differ for stars in different parts of the sky,
leading only to a general blurring of features.

\citet[][2011]{Klem08} present a more sophisticated analysis for
possible streams among nearby stars in the first two \RAVE\ data
releases, and \citet{KK11} have recently confirmed many of the
features in the $(U,V)$ plane from an analysis of \RAVE\ stars
selected to have smaller velocity uncertainties.

\begin{figure}
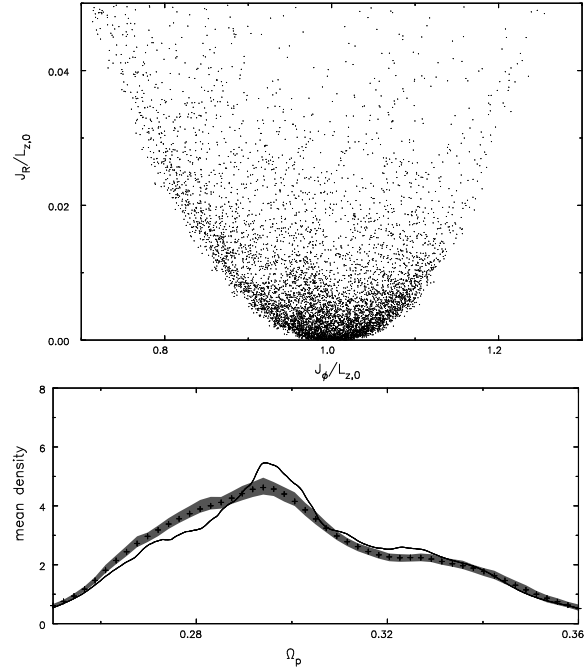

\includegraphics[width=.9\hsize,angle=0]{actplt.ps}
\includegraphics[width=.9\hsize,angle=0]{linint.ps}
\caption{The upper panel shows the distribution of actions estimated
  for the $6\,769$ stars of the combined sample.  The parabolic lower
  boundary is a selection effect, since only stars with larger radial
  oscillations can visit the solar neighbourhood if their guiding
  centre radii differ from $R_0$.  \hfil\break The lower panel shows
  the mean density of stars in the upper panel along resonance lines
  for the \ILR\ of $m=2$ disturbances having a range of pattern
  speeds.  The shaded region shows the 99\% confidence range from
  randomly resampled coordinates, while the solid line shows the same
  quantity from the selected sample.  The equivalent plot to the lower
  panel shows the same significant feature, except at a quite
  different frequency, when we assume trapping at either an \ILR\ or
  an \OLR\ of a 2-, 3- or 4-fold symmetric pattern.}
\label{actplt}
\end{figure}

\section{Analysis}
We compute action-angle variables \citep{BT08} for the in-plane
motions of the selected stars, as described in Paper I, using our best
estimates of the full phase-space coordinates for each star, corrected
for the motion of the local standard of rest \citep[\LSR][]{SBD}.
Monte Carlo simulation using the uncertainties in the input data
indicated that the median action uncertainties are $\sigma(J_R) =
0.0032$ and $\sigma(J_\phi) = 0.020$, while the median angle
uncertainties are $\sigma(w_R) = 0.18$ and $\sigma(w_\phi) = 0.024$
radians.  The dimensionless actions are scaled by $L_{z,0} = R_0V_0$,
where $R_0$ is the solar distance from the Galactic centre and $V_0$
is the orbital speed of the \LSR, in order to make them independent of
these two uncertain quantities.  The upper panel of Fig.~\ref{actplt}
shows the distribution of actions for the stars in the combined
sample.

Stars in resonance with a weak perturbation having $m$-fold rotational
symmetry and which rotates at the angular rate $\Omega_p$ have
unforced frequencies that obey the relation $m\Omega_\phi + l\Omega_R
= m\Omega_p$, where $l=0$ at corotation, and $l=\mp1$ at the \ILR\ and
\OLR\ respectively.  This frequency condition defines a line in action
space that slopes to smaller $J_\phi$ for increasing $J_R$ for all
$(l,m)$.

The lower panel shows the density of stars along \ILR\ ($l=-1$) lines
for $m=2$ disturbances having a range of pattern speeds.  As in Paper
I, we estimate the significance by comparison with equal size samples
of pseudo-stars but with radial coordinates and in-plane velocities
chosen independently in a bootstrap fashion from the distributions of
these three variables.  The shaded area encloses 99\% of the results
from these randomly resampled coordinates, while the solid line shows
the same quantity from the selected sample.  The peak value of the
solid curve is thus statistically highly significant and occurs for
$\Omega_p=0.294$ (in units of $V_0/R_0$), in excellent agreement with
the value (0.296) obtained from the completely independent
\GCS\ sample.

The excess at this frequency is caused by the overdense feature
visible in the upper panel slanting upward with negative slope from
near the point (1,0).  Unfortunately, this feature cannot be uniquely
attributed to an \ILR\ of an $m=2$ pattern, since the loci of
corotation and of both Lindblad resonances for $m\geq2$ also have
similar steep negative slopes in action space, as shown in the lower
panel of fig.~5 of Paper I.  Thus tests for {\em trapping} at any
resonance appear very similar to that shown in the lower panel of
Fig.~6, although the almost equally significant excess when other
resonances are considered clearly must imply quite different pattern
speeds.

Stars that are {\em scattered} at a resonance, on the other hand,
change both actions in such a way that the Jacobi constant is
conserved, which generally shifts a star in a direction in action
space that is not parallel to the resonance locus.  Only at the
\ILR\ does the scattering line in action-space have similar, though
not identical, slope to the resonance locus; scattering lines at
corotation are horizontal while they have positive slope at an \OLR.
Tests for an excess of stars along lines of positive slope in action
space, that would correspond to scattering at an \OLR, confirmed that
there are no significant features with this slope, as was also true
for the \GCS\ sample in Paper I.

\begin{figure}
\includegraphics[width=.9\hsize,angle=0]{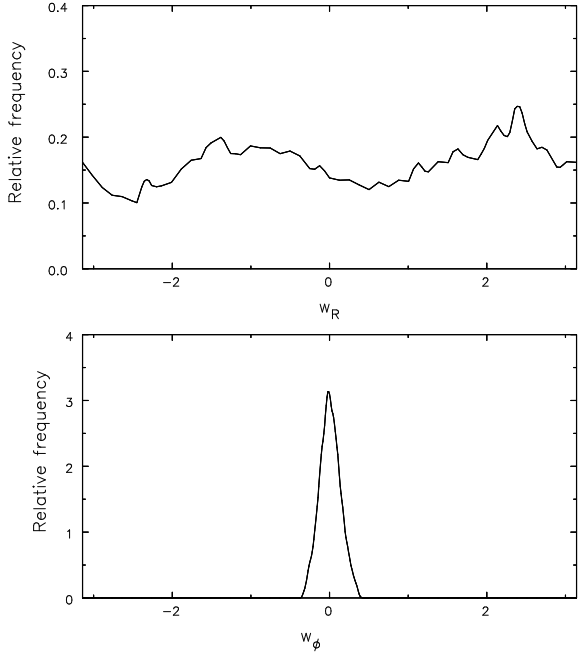}
\caption{Distributions of radial phase angles $w_R$ (upper panel)
  and of orbital phase angles $w_\phi$ (lower panel) for the combined
  sample.}
\label{angplt}
\end{figure}

Figure \ref{angplt} shows the distributions of the two angles
conjugate to the actions.  (The physical meaning of these variables is
explained in the Appendix.)  The orbital phase distribution, $w_\phi$,
is narrow because the guiding centres of all stars in the sample are
close to the Sun's azimuth in the Galaxy, which was arbitrarily chosen
to lie at $w_\phi = 0$.  The distribution of radial phases, $w_R$, is
non-uniform also, reflecting the substructure in phase space
(Fig.~\ref{contall}).

\begin{figure}
\includegraphics[width=\hsize,angle=0]{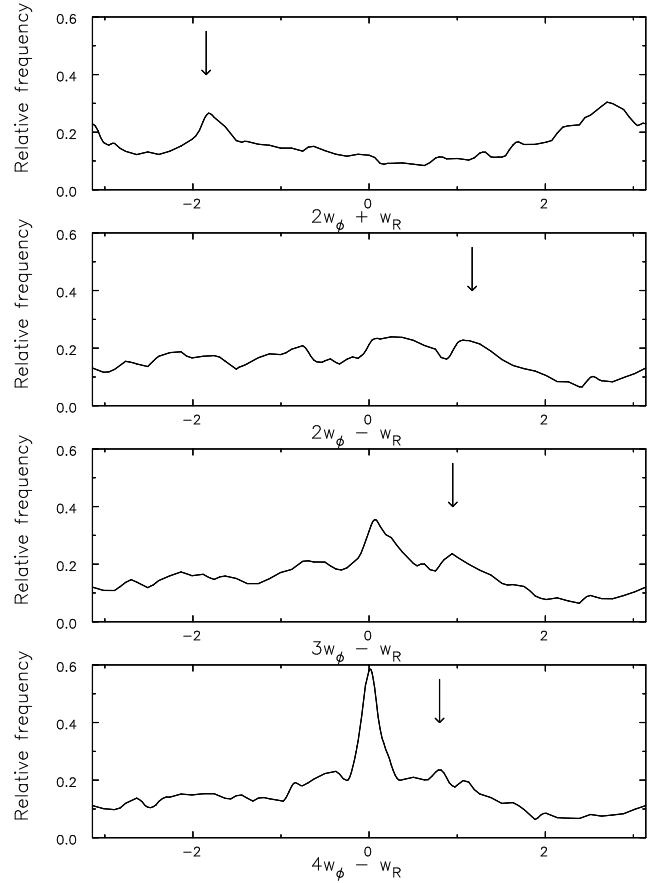}
\caption{Tests of the distributions of angles for trapping at an outer
  Lindblad resonance of an $m=2$ disturbance (upper panel) and
  respectively of an \ILR\ for $m=2$, $m=3$ and $m=4$ disturbances
  (lower three panels).}
\label{restest}
\end{figure}

As explained more fully in Paper I, a new peak that appears in any of
the distributions of simple linear combinations of these phase angles
may indicate a group of stars trapped in, or recently scattered by, a
resonance.  The appropriate combination is $mw_\phi + lw_R$.  This
test is insensitive to corotation, where $l = 0$, but a new
concentration of stars at some value of one of these combinations with
$|l|=1$ is an indicator of a Lindblad resonance.  \citet{McM11} has
shown that selection effects in any local sample, together with a
small amount of scatter about the expected constant value of $mw_\phi
+ lw_R$, conspire to frustrate this test.  In the light of his
finding, our more limited objective here is to show that the features
that appeared in the tests in Paper I have their counterparts in the
present sample.

Fig.~\ref{restest} shows the distributions of four combinations of the
angle variables of stars in the combined sample.  The top panel shows
the case for an \OLR\ for $m=2$, while the lower three panels are for
various \ILR s.  The overall shapes of the distributions differ in
detail from those found for \GCS\ stars in Paper I.  However, Sellwood
identified clear peaks in the \ILR\ cases for $m=2$ and $m=3$ (his
Fig.~7), and for $m=4$ in Fig.~\ref{restest4} of this paper.  These
features, at the abscissae marked by arrows, can also be identified in
Fig.~\ref{restest} in this independent sample.  Although the
significance of the peaks at the indicated abscissae is not high, they
do lie at the same phases as those in the \GCS\ sample.  The peak near
zero, that becomes more prominent as $m$ is increased, is a selection
effect that is explained in the Appendix.  A peak is also visible in
the \OLR\ distribution, in the top panel, near $2w_\phi+w_R \sim -1.8$
that also has a counterpart at the indicated position in the
corresponding distribution from the \GCS\ sample in Paper I, although
in that case the difference between the distributions of $2w_\phi+w_R$
and of $w_R$ alone was less pronounced.

The peaks in these new data are far from compelling, but they provide
a valuable confirmation of the far more significant peaks that were
found from the higher quality \GCS\ sample.  Measurement errors must
always smooth away features on the scale of the uncertainty and the
above estimates of the uncertainties suggest broadening on the scale
of $\sigma \sim 0.2$ radians.  In particular, it is reassuring that
the most significant peak in Fig.~\ref{actplt} lies at the same
frequency, and the peaks in the lower three panels of
Fig.~\ref{restest} lie at the same phases, as in the corresponding
Figures from the \GCS\ sample.  Taken together, this is strong
evidence in support of a Lindblad resonance from this completely
independent sample.

\section{Conclusions}
We have shown that the velocity-space substructure revealed by
\Hipp\ \citep{Dehn98} and the \GCS\ \citep{Nord04,HNA9} is also
present among the nearby ($d<200\;$pc) stars in the \RAVE\ and
\SDSS/Segue2 samples.  We find that the velocity space substructure
closely resembles that found for the independent sample of \GCS\ stars
even though the present sample is barely half the size and velocity
uncertainties are several times larger; most of the significant star
``streams'' of the \GCS\ have counterparts in the new sample, although
the precise velocities of the streams do not match perfectly.

Analysis of the action variables constructed from our best estimates
of the full phase space coordinates of each star in the sample reveals
an excess of stars along a resonance scattering trajectory in
action-space at a similar frequency to that found in Paper I
\citep{Sell10} for the \GCS\ sample.  While still statistically highly
significant, the feature in Fig.~6 stands out less clearly than in
Paper I because the sample is smaller and uncertainties are larger.
The evidence for resonant trapping in Fig~\ref{restest} is also weaker
than that found for the \GCS\ sample, but again reveals peaks at the
same phases.  We therefore consider this sample of stars to support
the evidence for a Lindblad resonance found in Paper I, although here
again the data do not constrain the multiplicity of the pattern.

While it is unfortunate that these data do not rule out trapping at an
\OLR\ as the cause of the Hyades stream, it is worth noting that
\citet{Dehn00} examined the effect on the local phase space due to the
\OLR\ of the bar and did not find any Hyades stream-like features.
Since the Hyades stream stars form the tongue in action-space that
stands out Fig.~3 of Paper I and our Fig.~\ref{actplt}, it seems to us
far more likely that the stream was created by an \ILR.
\citet{Sell94,Sell00} reported that features that extend upward
towards smaller $J_\phi \;(\equiv L_z)$ for increasing $J_R$ (or
$E_{\rm rand})$ are created by \ILR s in his simulations.  Since
scattering at an \OLR\ moves stars in action space in a direction that
is roughly perpendicular to the resonance locus, as discussed in \S3
above, stars do not move far before leaving the resonance.  Only in
the case of an \ILR\ do stars stay close to resonance as they get
pushed by the disturbance, thereby allowing stars to be moved from the
dense low-$J_R$ region up to higher $J_R$ where the overdensity stands
out.  Note that if the cause is a spiral (other types of disturbance
could be responsible), the \ILR\ of a bisymmetric spiral seems
unlikely, since it would place corotation unreasonably far out in the
disk, but an $m=4$, or perhaps even an $m=3$, spiral would seem more
likely, as noted in Paper I.  This speculation could be tested by data
from Gaia \citep{Perr01}, which will obtain phase space information
over a much more extensive region of the Galaxy.

\section*{Acknowledgments}
Correspondence with Paul McMillan has been most helpful.  We also
thank Ralph Sch\"onrich and an anonymous referee for constructive
comments on an earlier draft of this paper.

\appendix
\section[]{Selection effects}
\begin{figure}

\includegraphics[width=\hsize,angle=0]{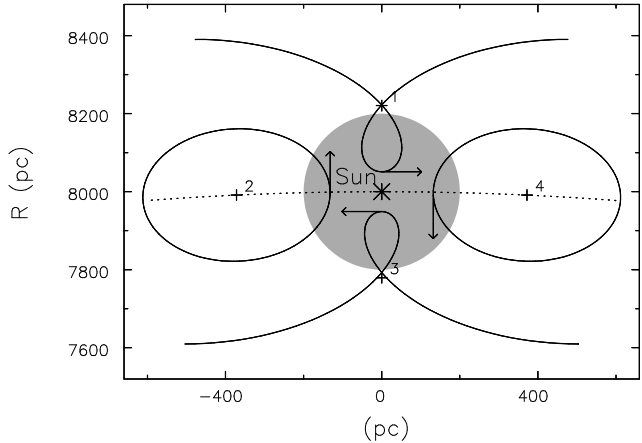}
\caption{Four possible orbits that pass through the vicinity of the
  Sun (asterisk), drawn in the frame that rotates at the angular rate
  of the \LSR.  The sense of rotation of the Galaxy is clockwise, in
  line with the usual convention.  The sample volume is the lightly
  shaded region, and the dotted line marks the solar circle.}
\label{draw}
\end{figure}

Both the present sample and the \GCS\ sample of Paper I were confined
to stars that happen to lie near the Sun at present.  Since stars move
on eccentric orbits, most stars in the sample are merely visiting the
solar vicinity from farther afield.

Fig.~\ref{draw} shows four possible orbits of stars near the Sun that
all happen to pass through the (shaded) survey volume.  The orbits are
drawn in the frame of the \LSR\ and have sufficiently small radial
excursions to approximate Lindblad epicycles.  All four orbits have
the same $J_R$ (epicycle size); those labelled 2 \& 4, with guiding
centres (plus symbols) lying on the solar circle, have the same
$J_\phi \;(=L_z)$ as does the Sun, while orbits 1 \& 3 have,
respectively, greater and smaller values of $J_\phi$.  Orbits 2 \& 4
are closed ellipses, because their guiding centres move at the same
angular rate as the frame, but the guiding centres of the other two
ellipses, which are marked at the point of the star's closest approach
to the Sun, drift relative to the \LSR.  For clarity, the orbits shown
do not loop around the Sun so that the vectors, which show peculiar
velocities relative to the \LSR, appear to correlate with the
direction of the star from the solar position.  In reality, stars
anywhere in the survey region can have peculiar velocities in any
direction.

\begin{figure}
\includegraphics[width=\hsize,angle=0]{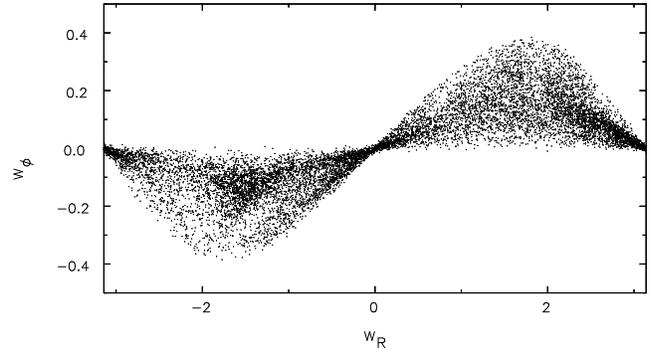}
\caption{The distribution of angles for \GCS\ stars.  Stars with
  radial action $>0.05L_{z,0}$ are excluded.  Note the difference in
  scales between the two axes.}
\label{angdist}
\end{figure}

The value of $w_\phi$ is the instantaneous Galactic azimuth of the
guiding centre, which \citet[][Paper I]{Sell10} chose to be zero on
the line connecting the Sun to the Galactic centre.  Thus $w_\phi>0$
for stars ahead of the Sun.  The $w_R$ variable is the phase of the
star around its epicycle, which increases with time in the sense shown
by the arrows.  Again \citet{Sell10} adopted $w_R=0$ for a star at its
apocentre and consequently $w_R=\pm\pi$ for a star at its pericentre.
Note that action-angle variables, unlike Lindblad's epicycles, can
also be used for more eccentric orbits, for which the guiding centres
generally orbit more slowly than the circular speed and the radial
oscillation is anharmonic.

Fig.~4 of Paper I showed the separate distributions of $w_R$ and
$w_\phi$ for stars in the \GCS\ sample having $J_R < 0.05L_{z,0}$.
Here, Fig.~\ref{angdist} shows the joint distribution of the same
stars in the space of both angles.  The azimuthal phase distribution
is confined to $|w_\phi|\la 0.4$ reflecting the limited spread in
Galactic azimuths of the guiding centres for stars that pass through
the survey volume.

The distribution of $w_\phi$ values for stars having $w_R=0$, such
as orbit 3 in Fig.~\ref{draw}, or $w_R=\pm\pi$ (orbit 1) is very
narrow because their guiding centres lie along the radius vector from
the Galactic centre to the Sun for all values of $J_R$.  Stars in the
first quadrant of Fig.~\ref{angdist} are inward moving stars (because
$w_R > 0$) with guiding centres ahead of the Sun ($w_\phi > 0$), such
as orbit 4.  Conversely, stars in the third quadrant ($w_\phi < 0$)
are outward moving stars having $w_R < 0$ (\eg\ orbit 2).  Thus the
selection criteria preclude stars from lying in the second and fourth
quadrants.  The spread of $w_\phi$ values for $w_R \sim \pm \pi/2$
reflects the spread in the sizes of the epicycles of the stars in the
sample, from which more eccentric orbits were eliminated.

\begin{figure}
\includegraphics[width=\hsize,angle=0]{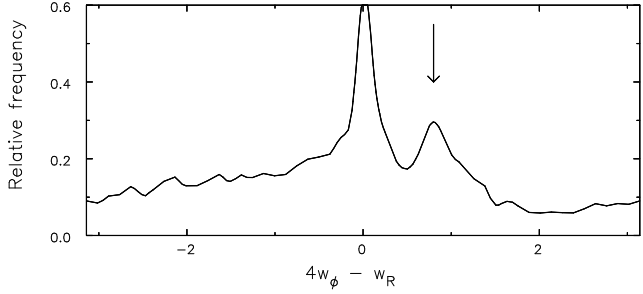}
\caption{The distribution of $4w_\phi - w_R$ for \GCS\ stars, which
  was not shown in Paper I.  As there, stars with radial action
  $>0.05L_{z,0}$ are excluded.}
\label{restest4}
\end{figure}

\citet{Sell10} argued that a concentration of stars having a nearly
constant value of $mw_\phi + lw_R$ would be an indicator of trapping
in a resonance for the selected values of $l$ and $m$.  Thus he
searched for an excess of stars lying along lines of fixed slope with
all possible intercepts in this plot, and reported the results in his
Fig.~7 of Paper I for the \GCS\ sample.  The lines for inner Lindblad
resonances ($l=-1$) have a positive slope in Fig.~\ref{angdist}.
Their slopes decrease as $m$ increases, causing them to become more
closely parallel to the distribution near $(w_R,w_\phi)=(0,0)$, giving
rise to a peak near $mw_\phi - w_R = 0$ as $m$ increases.  The
distributions shown in Fig.~7 of paper I for $m=2$ and $m=3$, and in
Fig.~\ref{restest4} of this paper, show the increasing prominence of
this feature, which is purely an artefact of the sample selection.  A
similar feature appears near $mw_\phi + w_R = \pm \pi$ when $l=1$,
since the lines have negative slope in these cases.

\citet{Sell10} drew attention to the peaks that lay away from
$mw_\phi-w_R=0$ for the cases of inner Lindblad resonances.  These
peaks arise in part from the concentration of \GCS\ stars near $w_R
\sim -1.5$, $w_\phi \sim -0.15$ in Fig~\ref{angdist}, but the clump is
visibly extended, which \citet{McM11} described as having a triangular
shape.  Since the excess lies in the third quadrant, it gives rise to
peaks in the distributions of $mw_\phi-w_R$ that shift closer to zero
as $m$ is increased.  Obviously the same excess gives rise to peaks
near $mw_\phi+w_R \ga -2$ for lines of negative slope in the cases of
an \OLR\ but, at least for the \GCS\ sample, they are not as striking
as those for the \ILR\ cases.

The case for an \ILR\ made by \citet{Sell10} did not rest solely on
this slight difference, however.  He attached far greater significance
to the fact that stars lying along a line of positive slope extending
through the clump were exactly those that lay along the resonant
scattering tongue in action space, as shown in Fig.~8 of Paper I.  As
this was not the case for stars lying on lines of negative slope,
Sellwood concluded that an \ILR\ was responsible.  Unfortunately,
\citet{McM11} found that his conclusion was compromised by selection
effects: if stars are not only in the vicinity of the Sun but are also
trapped in a resonance, McMillan was able to show that {\it both\/}
\ILR\ and \OLR\ models give rise to features in the angle distribution
(\ie, Fig.~\ref{angdist}) having positive slope.  Thus, although
scattering at a Lindblad resonance was clearly established, the test
in Paper I did not enable the character of the resonance to be
determined.

\label{lastpage}

\end{document}